\newcommand{\sss}{{\scriptscriptstyle }}

\def\eV{\hbox{eV}}

\def\roughly#1{\mathrel{\raise.3ex\hbox{$#1$\kern-.75em
\lower1ex\hbox{$\sim$}}}}
\def\lsim{\roughly<}
\def\gsim{\roughly>}
\def\pref#1{(\ref{#1})}

\documentstyle[prl,aps]{revtex}
\begin{document}
\input{psfig.tex}

\twocolumn[\hsize\textwidth\columnwidth\hsize\csname@twocolumnfalse%
\endcsname
\rightline{McGill-01/16, {\tt astro-ph/0107573}}
\vspace{3mm}

\draft
\title{Natural Quintessence and Large Extra Dimensions}

\author{A. Albrecht,${}^a$ C.P.~Burgess,${}^{b,c}$ F. Ravndal ${}^d$ 
and C. Skordis ${}^a$}

\address{${}^a$ Department of Physics, UC Davis,
1 Shields Avenue, Davis CA, USA 95616.\\
${}^b$ School of Natural Sciences, Institute for Advanced Study, 
Princeton NJ, USA 08540.\\
${}^c$ Physics Department, McGill University,
3600 University Street, Montr\'eal, Qu\'ebec, Canada H3A 2T8.\\
${}^d$ Institute of Physics, University of Oslo, N-0316 Oslo, Norway.}
\maketitle

\begin{abstract}
{
We examine the late-time (nucleosynthesis and later) cosmological
implications of brane-world scenarios having large (millimeter sized)
extra dimensions. In particular, recent proposals for understanding
why the extra dimensions are so large in these models indicate that
moduli like the radion appear (to four-dimensional observers) to be
extremely light, with a mass of order $10^{-33}$ eV, allowing them to
play the role of the light scalar of quintessence models. The
radion-as-quintessence solves a long-standing problem since its small
mass is technically natural, in that it is stable against radiative
corrections. Its challenges are to explain why such a light particle
has not been seen in precision tests of gravity, and why Newton's
constant has not appreciably evolved since nucleosynthesis. We find
the couplings suggested by stabilization models can provide
explanations for both of these questions. We identify the features
which must be required of any earlier epochs of cosmology in order for
these explanations to hold. 
}
\end{abstract}
\pacs{PACS numbers: }
]

%
%

 \section{Introduction} If current experimental indications are to be
believed, the universe is composed of islands of luminous material
embedded amongst darker baryonic matter, wrapped in the riddle of
nonbaryonic dark matter inside the enigma of dark energy.  Best fits
to cosmological models give the last three of these in the rough
relative proportions of few\% : 30\% : 70\% \cite{Cosmology}. It is
staggering that essentially nothing is known about {\it either} of the
two dominant components to the universe's present energy density.  

The very existence of dark energy presents serious puzzles.  The simplest
account of the cosmic acceleration seems to be a non-zero cosmological
constant $\Lambda$. However the energy scale of the cosmological
constant is problematic, requiring an energy density $\rho_{\rm
{\Lambda}} \approx v^4$ with $v \sim 10^{-3}$ eV.  The long-standing
cosmological constant problem\cite{Weinberg} stems from the fact that
vacuum fluctuations contribute to the cosmological constant, and
simple estimates calculated by assuming a Planck-mass cutoff give
$\Lambda$ a value $120$ orders of magnitude greater than is be
consistent with observations.  This discrepancy is reduced somewhat in
models where the natural cutoff is the supersymmetry breaking scale,
but even in that case there is a very serious problem.  

Many physicists
are attracted by the idea that the cosmological constant problem could
eventually be solved by some symmetry argument that would set
$\Lambda$ precisely to zero (although very interesting alternative views do
exist\cite{BanksFischler}. 
The observed cosmic acceleration creates problems  
with that line of reasoning, since it appears to indicate that $\Lambda$
actually is non-zero. One way out is to assume that $\Lambda$ really
is zero, and that something else is causing the cosmic
acceleration. This ``something else'' is usually called ``dark
energy'' or ``quintessence''\cite{FirstOne,Quintessence}.

For quintessence, typically one proposes the existence of a scalar
field whose mass is small enough for it to be evolving in a
cosmologically interesting way at
present. So far, essentially all quintessence models present the
following challenges or puzzles\cite{Naturalness}.  P1: Why don't
radiative corrections destabilize the fantastic hierarchy between $v$,
the scalar masses which are required ($m \sim v^2/M_p \sim 10^{-33}$
eV) and the other known scales of physics? (Only the first
quintessence model attempts to address this problem\cite{FirstOne}) P2:
Why isn't the very long-range force mediated by the new light scalar
detected in precision tests of gravity within the solar system?  

We here report on a class of quintessence models which is motivated by
recent efforts to understand the hierarchy problem, in which all
observed particles except gravitons (and their supersymmetric
partners) are trapped on a (3+1)-dimensional surface, or brane, in a
higher-dimensional `bulk' space \cite{ADD,ADD2}.  It turns out that
within this framework the above two puzzles turn out to be resolved in
a natural way. We believe ours is the only quintessence models to
successfully address {\em both} these particular challenges\footnote{
Ref. \cite{CormierHolman} discusses another type of quantum correction which
corrects classical rolling of the quintessence field off a local
maximum.  Those calculations do not apply to the type of potentials
considered here.}. 

A great virtue of the class of models which we explore is that they
generally arise in the low-energy limit of theories which were
originally proposed to solve the gauge hierarchy problem, without any
cosmological applications in mind. As we shall see, the generic
low-energy limit of these models looks like scalar-tensor gravity
\cite{STRefs}, with the four-dimensional scalars of relevance to
cosmology coming from some of the extra-dimensional modes of the
six-dimensional metric. We show here that although these models were
not proposed with cosmology in mind, the couplings they predict for
these scalars have following properties:

\begin{enumerate} 

\item {\it Good News:} They can
predict a late-time cosmology which provides a quintessence-style
description of the dark energy. And they do so with scalars whose
masses are extremely small, yet technically natural.

\item {\it Good News:} They can evade the constraints on the
 existence of very long-ranged forces because the couplings of
 ordinary matter to the the very light scalars are predicted to be
 {\it field dependent}, and so evolve over cosmological timescales.
 As has been observed elsewhere \cite{DamourNordtvedt}, once free to
 evolve, such couplings often like to evolve towards zero, implying
 acceptably small deviations from General Relativity within the solar
 system at the present epoch.

\item {\it Good News:} The
 expectation values of the scalar fields whose cosmology we explore
 determine the value of Newton's constant, and the success of Big-Bang
 nucleosynthesis (BBN) implies strong limits on how much this can have
 changed between then and now.  We find the evolution of Newton's
 constant changes very little over cosmologically long times, for a
 reasonably wide range of initial conditions when we enter the BBN
 epoch: Newton's constant tends not to change once the motion of the
 scalar field is either kinetic- or potential-dominated.

\item {\it
 Bad News:} The previous two cosmological successes are {\it not}
 completely generic, in the following sense. First, they depend on
 dimensionless couplings taking values which are of order $\epsilon
 \sim 1/50$ or so. (Although couplings of this order of magnitude
 naturally do arise in the models of interest, we require some modest
 coincidences in their values to ensure sufficiently small evolution
 in Newton's constant.)  Second, although many initial conditions do
 produce an acceptable cosmology, this success is not completely
 generic.  Attractor solutions exist which would cause too much
 variation in Newton's constant if they described the cosmology
 between BBN and today. Furthermore, obtaining the present-day value
 of Newton's constant requires adjusting the features of the scalar
 potential.
\end{enumerate} 

 We present our results here as an example of a viable cosmology,
 based on using the brane-world's extra-dimensional `radion' mode as a
 quintessence field. Our ability to do so may itself be considered
 something of a surprise, since extra-dimensional aficionados had
 earlier examined the radion as a quintessence candidate, and found it
 wanting. Our conclusion here differs, because of our recognition that
 the radion couplings tend to be field dependent, in a way we make
 precise below.
 
 We regard this work as a first exploration of the conditions which a
 successful cosmology must assume. We believe our results to be
 sufficiently promising to justify further exploration of this class
 of models, to see to what extent the unattractive features can be
 improved upon, and to ascertain what observations can test whether
 the universe indeed proceeds as we describe here.
 
 Our presentation is organized as follows. In the next section we
 briefly describe the brane models we use, and their motivations and
 problems. The low-energy action relevant for cosmology is also
 presented in this section. Section 3 describes some of the
 cosmological solutions of the theory, focusing in particular on
 analytic expressions on which intuition can be based. More
 complicated, and realistic, solutions are then described
 numerically. We close in Section 4 with a discussion of our results.
 
 \section{The Model}
 
 Our point of departure is the brane-world scenario in which all
 observed particles except the graviton are trapped on a
 (3+1)-dimensional surface, or brane, in a higher-dimensional `bulk'
 space. In particular, we imagine the scale of physics on the brane is
 $M_b \sim 1$ TeV and the bulk space is approximately six-dimensional,
 comprising the usual four, plus two which are compact but with radius
 as large as $1/r \sim 10^{-3}$ eV \cite{ADD,ADD2}.  The large ratio
 $M_b r \sim 10^{15}$ is phenomenologically required to reproduce an
 acceptably weak gravitational coupling in four dimensions, $M_p = (8
 \pi G)^{-1/2} \sim M_b^2 r \sim 10^{18}$ GeV.
 
 These kinds of models face two main difficulties, one theoretical and
 one phenomenological.  Theoretically, these models trade the large
 hierarchy $M_b/M_p = 10^{-15}$ for the hierarchy $M_b r = 10^{15}$,
 but an explanation of this last value ultimately must be
 provided. Recently progress towards providing this explanation has
 been made, by proposing energetic reasons why the extra-dimensional
 radii might want to prefer to take large values
 \cite{Other6d,ABRS}. Our quintessence model does not rely on the
 details of these mechanisms, but our choice for the scalar field
 couplings is strongly motivated by the mechanism of ref.~\cite{ABRS}.
 
 Phenomenologically, there are strong upper limits on how large $M_b
 r$ can be. These limits come from supernova \cite{SNLimits,LED} and
 cosmological observations \cite{ADDCosmo} which preclude the
 existence of Kaluza-Klein (KK) partners of the bulk-space graviton
 and its superpartners. We believe the present-day observational
 constraints on KK modes can be accommodated if only two dimensions
 have $r$ as large as $(10^{-3} \eV)^{-1}$, but prevent $r$ from being
 taken any larger.  Although the cosmological constraints generally
 impose stronger upper limits on $r$ they depend more on the details
 of the assumed cosmology, and can be evaded under certain conditions
 (which we describe in more detail later).
 
 \subsection{The Low-Energy Action}
 
 We start by sketching the brane world scenario, and its low energy
 action.  Our discussion closely follows that of ref.~\cite{ABRS}.
 
 The couplings of the brane-world model relevant to cosmology are
 those of the six-dimensional metric, ${\cal G}_{\sss MN}$. (Six
 dimensions are required in order to allow the extra dimensions to
 have radii as large as we need.)  The leading terms in the derivative
 expansion for the six-dimensional action have the form $S = S_B +
 S_b$, where the Bulk action describing the metric degrees of freedom
 is: 
\begin{equation} 
\label{Lagr} S_B = \- \,{M_b^4 \over 2} \int
 d^4x d^2y \sqrt{{\cal G}} \; {\cal R} + \cdots, 
\end{equation} 
 where ${\cal R}$ denotes the scalar curvature built from the
 six-dimensional metric, and we write no cosmological constant for the
 bulk.\footnote{This can be ensured by symmetries, such as in a
 supersymmetric compactification \cite{SezSal}.  Like everyone else we
 assume the vanishing of the effective four-dimensional cosmological
 constant, and our ideas do not add to the understanding of why this
 should be so.}
 
 The dependence of the brane action on its various matter fields is
 not crucial for our purposes, but what is important is that it
 depends only on the metric evaluated at the brane's position: 
 \begin{equation}
 S_b = \int d^4x \sqrt{g} {\cal
 L}_b(g_{\mu\nu},\dots), 
\end{equation} 
where the ellipses denote
 all of the other fields on which ${\cal L}_b$ could depend.
 
 The tree-level dimensional reduction of this action using the metric
  \begin{equation} 
\label{MetricAnsatz} {\cal G}_{\sss MN} =
 \pmatrix{ \hat{g}_{\mu\nu}(x) & 0 \cr 0 & \rho^2(x) \, h_{mn}(y)
 \cr}, \end{equation} 
 with $\rho = M_b \, r$, gives the leading
 contribution to the effective four-dimensional Lagrangian: 
 \begin{equation} 
\label{KineticTerms1} {\cal L}_{\rm kin} = - {M_b^4
 r^2 \over 2} \; \sqrt{\hat{g}} \left[R(\hat{g}) - 2 \left({\partial r
 \over r} \right)^2 + { R(h) \over M_b^2 r^2} \right], 
\end{equation}
 where we have adopted the conventional normalization $M_b^2 \int
 d^2y \; \sqrt{h} = 1$.  Here $R(\hat{g})$ and $R(h)$ denote the
 curvature scalars computed from the metrics $\hat{g}_{\mu\nu}$ and
 $h_{mn}$, respectively.
 
 Quantum corrections modify this action in several important
 ways. First, the kinetic terms can become renormalized, introducing
 the possibility that their coefficients can acquire a dependence on
 $r$. Second, terms not written in eq.~\pref{KineticTerms1} can be
 generated, including in particular a no-derivative potential,
 $U(r)$. At large $r$, $U(r)$ can typically be expanded in powers of
 $1/r$, with the leading term falling off like $1/r^p$, for some $p\ge
 0$.  For instance, for toroidal compactifications, the leading
 contribution can arise due to the Casimir energy, which produces the
 potential $U(r) = U_0/r^4 + \cdots$. 
 
 We are in this way led to consider a scalar-tensor theory
 \cite{STRefs,DamourNordtvedt,Will}, for the couplings of $r$ and the
 four-dimensional metric, which we write as follows:
 \begin{eqnarray}
 \label{STAction}
 {{\cal L} \over \sqrt{-\hat{g}}} &=& - \frac{M_b^4 r^2}{2} \, \left
 [ A (r)R(\hat{g}) - 2 B(r) 
 \left( {\partial r \over r} \right)^2 \right] \nonumber\\
 && \qquad \qquad - U(r) - {{\cal L}_m(\hat{g}) \over \sqrt{-\hat{g}}}.
 \end{eqnarray}
 
 As just described, the tree-level dimensional reduction of the 6d Einstein-Hilbert 
 action predicts $A = B = 1$ and $U = 0$, and with these choices 
 eq.~\pref{STAction} describes a Jordan-Brans-Dicke theory with coupling parameter 
 $\omega = - 1/2$ \cite{JBD}. Notice that the matter Lagrangian, ${\cal L}_m$, 
 describes the couplings of all of the known particles, and only
 depends on $\hat{g}_{\mu\nu}$ 
 but not on $r$. (Ordinary matter does not couple directly to $r$ in the brane-world
 picture because it is trapped on the brane, and does not carry stress energy in 
 the extra dimensions.) 
 
 Notice that if the scale of $U(r)$ is completely set by $r$, as is true for
 the cases discussed in \cite{ABRS}, then the mass which is predicted for this
 scalar is incredibly small: of order $m \sim v^2/M_p \sim 10^{-33}$ eV. 
 It is the existence of such a small mass which allows $r$ to potentially play
 a cosmologically interesting role right up to the present epoch, and so to
 be a candidate for a quintessence field.
 
 Since the predictions for $A,B$ and $U$ depend crucially on the
 approximation  made in their derivation, in general we expect them to
 be given by more complicated functions. This turns out to be true,
 in particular, within the hierarchy-generating  mechanism described
 in ref.~\cite{ABRS}. Since we use this proposal  to motivate our
 choices for  these functions, we briefly pause here  to summarize its
 implications.  
 
 The proposal supposes that the spectrum of bulk states includes
 a six-dimensional scalar, $\phi$, which counts amongst its interactions a 
 renormalizable cubic self-coupling, $g \phi^3$. Since the coupling $g$ of 
 this interaction is dimensionless, it can introduce corrections to
 the low-energy  action which are logarithmic in $r$. Since all other
 couplings have negative mass  dimension in 6d, their quantum
 contributions are suppressed by powers of $1/r$,  implying the
 coupling $g$ can dominate at large $r$.  
 
 The bottom line of this scenario is that the dominant corrections to $A,B$ and $U$
 are of the form $1 + c_1 \, \alpha(r) + c_2 \, \alpha^2(r) + \cdots$,
 where the running loop-counting parameter is given by $\alpha(r) =
 \alpha_0 + \alpha^2_0 \log(r/r_0)  + \cdots$, with $\alpha_0 =
 g_0^2/(4 \pi)^3$. In this scenario the potential  $U(r)$ is plausibly 
 minimized for $\alpha(r) \sim 1$, implying a minimum  for which
 there is the desired large hierarchy: $M_b r \sim \exp[+ 1/\alpha_0]
 \gg 1$.  It is generic in this scenario that the quantum corrections
 to the other low-energy functions, $A$ and$B$, are also likely to be large. 
 
 For illustrative purposes when exploring cosmology, we choose the functions $A,B$
 and $U$ which are suggested by this picture in the perturbative
 regime, for which $A \approx 1 + a \log(M_b r)$, $B \approx 1 + b \log(M_b r)$ and
 $U \approx \left({ U_0 \over r^4} \right) \left[ 1 + c \log(M_b r) \right]$. Here
 $a,b$ and $c$ are small, since they are proportional to the
 dimensionless coupling constant, $\epsilon \sim g^2/(4\pi)^3$, and
 higher orders in $\epsilon$ give higher  powers of $\log(M_b r)$
 although these have not been written explicitly. In the spirit of this picture 
 we suppose $\epsilon$ to be of order $1/50$, as required in order to ensure 
 that $U$ is minimized when $M_b r \sim 10^{15}$. For cosmological
 purposes, we shall be interested in the regime for which $\epsilon \log(M_b r) 
 \lsim 1$ is negligible, but $\epsilon \log^2(M_b r)$ is not. 
 We take $U_0$ to be generic in size: $U_0 \sim O(M_b^4)$.  
 
 \subsection{Naturalness}
 
 Since naturalness is one of our motivations, we briefly summarize the
 arguments presented  in ref.~\cite{ABRS} as to why radiative
 corrections do not destabilize the scalar  mass. The argument takes a
 different form for the contribution due to the  integrating out of
 scales above, or below, $v \sim 1/r \sim 10^{-3}$ eV.  
 
 For scales below $v$, the effective theory consists of a four-dimensional 
 scalar-tensor theory coupled to itself, and to ordinary matter with couplings, 
 $\kappa$, of gravitational strength: $\kappa \sim 1/M_p$. Radiative corrections
 to the scalar mass within this effective theory are then generically acceptably small,
 being of order $\delta m^2 \sim \kappa^2 \Lambda^4$, with ultraviolet
 cutoff $\Lambda \sim v$.   These are therefore the same size as the
 tree-level mass, as is required if they  are to be technically natural.
 
 Radiative corrections to the masses coming from higher scales, between $v$ and $M_b$,
 are the dangerous ones. Generically, these are too large and it is the integrating
 out of these scales which introduces the naturalness problem to most quintessence
 models. Within the framework of ref.~\cite{ABRS}, the contributions from these
 scales are suppressed by supersymmetry, which is proposed to be broken at the scale
 $M_b$ on the brane on which observable particles live. (This ensures acceptably large
 splittings between the masses of ordinary particles and their superpartners.)
 
 Supersymmetry suppresses corrections to the radion mass because it relates $r$ to
 the four dimensional graviton, which is massless. Indeed, above the scale $v$ the bulk 
 space is really six-dimensional, and the radion is simply a component of the 
 six-dimensional metric. As such, its properties are tied to those of the 
 four-dimensional metric by the supersymmetry of the bulk space. Since the metric 
 and its superpartners only couple to the supersymmetry breaking with gravitational 
 strength, supersymmetry-breaking effects in the bulk sector are suppressed to be 
 of order $M_b^2/M_p \sim 1/r \sim v$.
 
 Since the scalar potential for the radion cannot arise until supersymmetry breaks, it
 is naturally of order $v^4 F(\varphi)$ where $\varphi$ is the dimensionless component
 of the six-dimensional metric which defines the radion. $\varphi$'s kinetic term,
 on the other hand, comes from the 6d Einstein-Hilbert action, and so is allowed by
 unbroken supersymmetry. At tree level it is of order $M_p^2 (\partial \varphi)^2$. 
 and the coefficient $M_p^2$ is much larger than the contributions
 produced by integrating  out physics at scales smaller than $M_b \ll
 M_p$. Again we are led on dimensional  grounds to a mass which is of
 order $m \sim v^2/M_p$.  
 
 \subsection{Cosmological Equations}
 
 The cosmology of the radion is most easily described by passing to
 the Einstein frame, for  which the radion and metric are canonically
 normalized. For the choices of $A = 1 + a \log(M_b r)$ and $B = 1 + b
 \log(M_b r)$ the canonically normalized variables are: 
 \begin{equation}
 \hat{g}_{\mu\nu} = g_{\mu\nu}/[(M_b r)^2 A^2], \quad \hbox{and} 
 \quad \log(M_b r) = \hat{a} \hat\chi (1 - \hat{b} \hat\chi),
 \end{equation}
 with $\hat{a} = (1 - 3a/4)/2$, $\hat{b} = (a-b)/16$ and $\hat\chi =
 \chi/M_b$. This
 choice of variables is called the Einstein frame, to distinguish it from
 the Jordan frame, as defined by eq.~\pref{STAction}.\footnote{A note 
 on numerics: In the Jordan frame $M_p  \sim M_b^2 r$, ordinary
 particle masses are $m_p \sim M_b$ and Kaluza-Klein  masses are
 $m_{\rm kk} \sim 1/r$, while in the Einstein frame $M_p \sim M_b$,
 $m_p \sim 1/r$ and $m_{\rm kk} \sim 1/(M_b r^2)$. Clearly, ratios of
 masses  like $m_{\rm kk}/m_p = 1/(M_b r)$ are the same in either
 frame, but  $M_b=1$ corresponds to using TeV units in the Jordan
 frame, but Planck  units in the Einstein frame.} 
 
 In the Einstein frame the action, eq.~\pref{STAction}, becomes
 \begin{equation}
 \label{EFAction}
 {{\cal L} \over \sqrt{g}} = - \, {M_b^2 \over 2} \, R - \frac12 \,
 (\partial \chi)^2   - V(\chi) - {{\cal L}_m(g,\chi) \over \sqrt{g}}, 
 \end{equation}
 where 
 \begin{eqnarray}
 \label{EFPot}
 V(\chi) &=& {U[r(\chi)]\over (M_b r)^4 A^2} \\
 &=& V_0 \Bigl[ 1 + (v-2a) \hat{a} \hat\chi +...\Bigr]
 \exp[- \lambda_0 \hat\chi (1 - \hat{b} \hat\chi) ] ,\nonumber
 \end{eqnarray}
 and $\lambda_0 = 8 \hat{a}$. (Even though $\hat{b} = O(\epsilon)$ we
 do not expand the second term in the exponent in this expression because
 such an expansion presupposes $\epsilon \hat\chi^2 \ll 1$, which is not
 true for our applications.)
 
 For cosmological applications we take $ds^2 = -dt^2 + a^2 d\vec{x}^2$
 in the Einstein  frame, in which case the Lagrangian \pref{EFAction}
 implies the usual  Friedman equations: 
 \begin{equation}
 \label{FEq}
 3 H^2 M_b^2 = \rho, \qquad 2\dot{H} M_b^2 = - (\rho + p),
 \end{equation}
 where both the energy density, $\rho$, and pressure, $p$, receive
 contributions from $\chi$ and the radiation and matter sectors --
 $\rho = \rho_r +  \rho_m + \rho_\chi$ and   $p = p_r + p_m + p_\chi$,
 where $p_r  \approx \frac13 \, \rho_r$ and $p_m \approx 0$.  
 
 In the Einstein frame the matter contributions also depend explicitly
 on $\chi$: 
 \begin{equation}
 \rho_m = {\rho_{m0} \over a^3 (M_b r)}, \quad
 \rho_r = {\rho_{r0}\over a^4}, 
 \end{equation}
 where the quantities $\rho_{m0}$ and $\rho_{r0}$, may be related to the
 density of matter and radiation at the present epoch (which we choose to
 be $a=1$). 
 
 Eqs.~\pref{FEq} must be supplemented with the evolution
 equation for $\chi$, which is:
 \begin{equation}
 \label{ChiEq}
 \ddot\chi + 3 H \dot\chi + V'(\chi) + {\eta(\chi) \over
 M_b} {T^\mu}_\mu =0,
 \end{equation}
 where $T_{\mu\nu}$ is the stress energy of the radiation and matter,
 and $\eta(\chi) = -\hat{a}(1 + a/2 - 2 \hat{b} \hat\chi)$. The successful
 comparison of General Relativity with precision tests of gravity within 
 the solar system require $\eta^2(\chi_0) \lsim 10^{-3}$
 when evaluated at the present epoch, $\chi = \chi_0$ \cite{DamourNordtvedt,Will}.
 Notice that since radiation satisfies ${T^\mu}_\mu = 0$, only matter 
 contributes to this last coupling, which is proportional to
 ${T^\mu}_\mu = \rho_m$.
 
 \section{Cosmology}
 
 We now show that cosmologically reasonable solutions may be found 
 to eqs.~\pref{FEq} and \pref{ChiEq}, which satisfy all of the bounds
 to which the theory must be subject. Since our goal is to demonstrate
 the existence and illustrate some of the features of these
 solutions, our analysis is not systematic and should be regarded as
 only preliminary explorations of models of this type. 
 
 \subsection{Constraints}
 
 Our main assumption is that the universe enters into an effectively four-dimensional
 radiation-dominated evolution at some temperature $T_*$ higher than a few MeV, 
 and so before Big Bang Nucleosynthesis (BBN). This assumption is a prerequisite
 for describing cosmology in terms of four-dimensional fields, and seems likely
 to be required if the success of standard BBN is not to be forsaken.
 
 There is a limit to how high $T_*$ may be chosen, since the four-dimensional
 picture requires the extra-dimensional KK modes to be cold. If $T_*$ is too
 high, then these KK modes can be thermally excited too efficiently through
 their couplings with ordinary particles. If these modes are too
 abundant they can overclose  
 the universe or decay too frequently into photons after recombination, and so 
 contribute unacceptably to the diffuse gamma ray background \cite{ADD,ADDCosmo}.
 If the KK modes all decay into observable daughters, then these constraints
 require $T_*$ not to be larger than the TeV scale, and can preclude having
 $1/r \sim 10^{-3}$ eV. 
 
 As the authors of refs.~\cite{ADD,ADDCosmo} themselves point out,
 these constraints may be evaded, depending on the cosmology which is
 assumed at earlier times in the universe's history. For instance,
 contributions to the diffuse gamma ray background is suppressed if
 the KK modes decay quickly enough into unobserved modes to be no
 longer present after the recombination epoch, when photons decouple
 from matter. Such fast decays can be arranged, for instance, if the
 KK modes decay more frequently into particles on other branes. Such
 decays may, in fact, be likely in cosmological scenarios for which
 the very early universe is dominated by a gas of branes and
 antibranes \cite{BraneGas}. (Another escape is possible if the
 extra-dimensional space is hyperbolic instead of toroidal, with its
 volume much greater than the appropriate power of its radius of
 curvature: $V \gg r_c^2$ \cite{Hyperbolic}. In this case KK modes can
 be very heavy, $m_{\rm kk} \sim 1/r_c$, even though the hierarchy
 problem is solved by having $V$ large.)  Although our present
 interest is for $T_* < 1$ TeV, we have no difficulty imagining some
 variation on these themes being invoked if scenarios requiring $T_* >
 1$ TeV were of interest. 
 
 Our second major constraint is the requirement that the quintessence
 field, $\chi$, not destroy the successes of BBN. This constraint
 arises because the success of BBN may be regarded as indicating that
 the hierarchy $M_w/M_p \propto (G_N/G_F)^{-1/2}$ is not much
 different at the epoch of nucleosynthesis than it is today. A
 sufficient condition for this to be true is to ensure that $M_b r$
 not to have evolved by more than about 10\% of its present value
 between BBN and the present epoch. This constraint is much stronger
 than what is required by garden-variety quintessence models. It is
 very strong because it requires that the field $r$ cannot have rolled
 significantly over cosmologically long times. A special property of
 the cosmological solutions we shall examine is that they often have
 the property that $r$ becomes fixed over cosmologically large
 timescales. Motivated by the BBN constraints we are particularly
 interested in those cosmological solutions for which $r$ does not
 strongly vary after BBN. We shall see that such solutions can arise
 for reasonably generic initial conditions for the functions $V(\chi)$
 and $\eta(\chi)$ we are using. 
 
 Next, we must demand agreement with present-day observations, which come
 in several important types. First, precision measurements of
 gravitational forces indicate that the coupling $\eta(\chi)$ must be
 $\lsim 0.03$ in order not to be in conflict with the success of
 General Relativity. Second, if $\chi$ is to provide an explanation of
 the dark energy, then its energy density must now be beginning to
 dominate that of matter, with $w = p_\chi/\rho_\chi < - \frac13$ if
 this energy density is to be currently accelerating the universe.  
 
 Finally, we require $\chi$ not to ruin the success of scale-invariant
 fluctuations in describing present-day observations of the cosmic
 microwave background radiation (CMB). Since a detailed analysis of
 the CMB in these models is beyond the scope of the present paper, we
 instead apply a poor-man's bound by requiring cosmological properties
 not to be too different from the standard picture near the epochs of
 radiation/matter crossover and recombination.
 
 \subsection{Approximate Cosmological Solutions} Before laying out a
 cosmology which satisfies the above constraints, it is worth building
 intuition by outlining several approximate analytical solutions to
 the cosmological equations which turn out to control the features
 which are seen in the numerical results. Since the microphysics
 motivates using the potential: $V(\chi) = V_p \; e^{- \lambda \chi}$,
 with $V_p(\chi)$, $\lambda(\chi)$ and the coupling $\eta(\chi)$
 varying much more slowly as (polynomial) functions of $\chi$ than
 does the exponential $e^{\lambda \chi}$, we explore solutions to the
 equations when $\lambda, V_p$ and $\eta$ are all constants. The
 cosmological evolution of such fields in the limit $\eta \to 0$ has
 been much studied \cite{ExpPots}, and our results go over to these in
 the appropriate limits.
 
 There are two types of solutions in this limit which are of principal
 interest. These solutions may be classified according to which the
 terms dominate in the $\chi$ equation, eq.~\pref{ChiEq}.
 \subsubsection{Scaling Solutions}
 
 The first class of approximate solutions we consider are the scaling,
 or tracker, solutions, of which there are two different types. These
 solutions are characterized by having all energy densities scaling as
 a power of the scale factor, and either: ($i$) $\ddot\chi \sim 3 H
 \dot\chi \sim V' \gg \eta \rho_m$ or, ($ii$) $\ddot\chi \sim 3 H
 \dot\chi \sim \eta \rho_m \gg V'$. Both types of solutions can arise
 regardless of whether it is the scalar, radiation or matter which is
 dominating the evolution of the universe.
 
 Assume, therefore, the scale factor scales with time as $a/a_0 =
 (t/t_0)^k$, and so $H = \dot a/a = k/t$. From the Friedman equation,
 $H^2 \propto \rho$, we see that if the dominant energy density scales
 as $\rho \propto a^{-n}$, then we must have $k = 2/n$.
 
 \medskip\noindent ($i$) $\ddot\chi \sim 3 H \dot\chi \sim V' \gg \eta
 \rho_m$: In this regime we assume 
\begin{equation} V \propto
 e^{-\lambda \hat\chi} \propto {1\over r^8} \propto {1 \over a^s},
 \end{equation} 
and we find from the $\chi$ field equation,
 eq.~\pref{ChiEq}, that $s = n$.
 
 If radiation or matter dominate the energy density, then $n=3$ or
 $n=4$. In this case we find solutions for which $K = \frac12
 \dot\chi^2 \propto V(\chi) \propto {1/ a^n}$, and so $r \propto
 a^{n/8}$. Eq.~\pref{ChiEq} also implies that these scaling solutions
 are trackers in the sense that $\rho_\chi$ tracks at precisely the
 fraction $n/\lambda^2$ of the dominant energy density as the universe
 expands. Solutions of this type only are possible if $\lambda^2 >
 n$. For $\lambda^2 < n$ the scalar energy density tends towards
 scalar domination of the universal expansion.
 
 If it is $\chi$ itself which dominates the energy density, then $k =
 2/\lambda^2$ and a scaling solution of the type we seek exists so
 long as $\lambda^2 < 6$.  In this case both $K$ and $V$ scale as
 $1/a^{\lambda^2}$, and so $r \propto a^{\lambda^2/8}$.
 
 These are the tracker solutions which are familiar from standard
 discussions of scalars evolving with exponential potentials
 \cite{ExpPots,scalarreview}.  Clearly these tracker solutions imply a
 relatively rapid variation of $r$ as the universe evolves.
 
 \medskip\noindent ($ii$) $\ddot\chi \sim 3 H \dot\chi \sim \eta
 \rho_m \gg V'$: We next consider the case where the $\eta
 {T^\mu}_\mu$ term dominates the scalar potential term in
 eq.~\pref{ChiEq}. Notice that this can hold even if both $\rho_r$ and
 $\rho_\chi$ are much larger than $\rho_m$, so the tracker solution we
 here obtain can be relevant even when nonrelativistic matter makes up
 only a very small part of the universe's total energy budget.
 
 Under these assumptions, because the last term of eq.~\pref{ChiEq}
 dominates the potential term, we ask $\rho_m$ to scale like
 $\ddot\chi$ and $H \dot \chi$: 
\begin{equation} \rho_m \propto {1
 \over a^3 r} \propto {1 \over a^s}, \end{equation} 
which, when
 inserted into eq.~\pref{ChiEq} implies $s=n$ and so $K \propto \rho_m
 \propto \rho \propto 1/a^n$.  As usual, if the universe is radiation
 or matter dominated then $n=4$ or $n=3$.  (Interestingly,
 both matter {\it and} radiation fall as $1/a^4$ within this 
 kind of tracker solution in a radiation-dominated universe.) 
 Alternatively, $K \propto
 \rho_m \propto \rho \propto 1/a^{\lambda^2}$, if the universe is
 $\chi$-dominated.
 
 Although the scaling of scalar kinetic energy is the same for tracker
 solutions of type $(i)$ and type $(ii)$, they differ markedly in the
 scaling which they imply for $r$. In the present solution the scaling
 of $r$ vs $a$ follows from the known scaling of $\rho_m$. For
 radiation/matter domination, where $\rho \propto 1/a^n$ we find $r
 a^3 \propto a^n$ and so $r \propto a^{n-3}$. Clearly $r \propto a$
 for radiation domination, but $r$ only varies logarithmically during matter
 domination! During a scalar-dominated phase, on the other hand, we
 have $r \propto a^{\lambda^2 - 3}$. Again, $r$ does not vary if
 $\lambda^2 = 3$.
 
 What is striking about these solutions where $r$ changes so slowly is
 that this happens in spite of its not sitting at the minimum of
 the potential. It happens, even though the radion's 
 kinetic energy dominates its potential energy, because the 
 amount of roll is limited by the Hubble damping generated 
 by the dominant energy component of
 the universe.

 \subsubsection{Transient Solutions}
 
 What is perhaps even more striking than the trackers for which $r$
 slowly varies is the behavior of $r$ during the transient evolution
 as the tracker -- which is an attractor -- is approached. Since the
 approach to the tracker solution may be approached from initial
 conditions for which $K \gg V$ or $K \ll V$, we consider each of
 these in turn.
 
 \medskip\noindent {\it $K \ll V$ (Slow Roll):} If $K\ll V$ then it is
 a good approximation to neglect $\ddot \chi$ in eq.~\pref{ChiEq}
 relative to the other terms. If $\eta \rho_m$ is also negligible
 relative to $V'$, then the evolution is described by $3 H \dot\chi +
 V' = 0$, which has as solutions: 
\begin{equation} \label{SlowRoll}
 V(a) = {2 V_{\rm exit} a^p_{\rm exit} \over a^p + a_{\rm exit}^p} ,
 \end{equation} 
where $p = n$ for radiation ($n=4$) and matter
 ($n=3$) domination, and $p = \lambda^2$ for scalar domination. Along
 this trajectory the kinetic energy grows as $K /V = \lambda^2 V/(6
 \rho)$.
 
 This solution implies $V \approx V_{\rm exit}$ is a constant (and so
 also must be $r$) until $a \sim a_{\rm exit}$, after which point the
 solution joins the tracker solution, for which $\rho_\chi \propto
 1/a^p$ and $\rho_\chi/\rho = p/\lambda^2$.
 
 \medskip\noindent {\it $K \gg V$ (Kinetic Domination):} If $K$
 dominates the scalar evolution, then $V'$ and $\eta \rho_m$ may be
 dropped relative to $\ddot\chi$ and $3H \dot\chi$. In this case the
 $\chi$ equation may again be solved analytically, with solution:
 $\chi = C_1 - C_2 \left( {a_0 / a} \right)^{3-1/p}$, where $C_1, C_2$
 are integration constants with $C_2$ related to the initial velocity
 by $C_2 = (3-1/p) a_0 (d\chi/da)_0$.  Here, as usual, $p = 2/n$ for
 radiation/matter domination, and $p = 2/\lambda^2$ for scalar
 domination.
 
 This solution implies the kinetic energy, $K = \frac12 \dot\chi^2$
 varies as: 
\begin{equation} \label{KinDomRoll} K(a) = K_0 \left( {
 a_0\over a} \right)^6, \end{equation} 
 regardless of the value of
 $p$.
 
 The evolution of the potential energy is more interesting in these
 solutions. Consider first the case of radiation/matter domination,
 for which $p = 2/n$ so 
\begin{equation} V(a) = V_0 \exp\left\{ c
 \left[ \left({a_0 \over a}\right)^{3-n/2} - 1 \right] \right\},
 \end{equation} 
with $c = (3 - 1/p) \lambda a_0 (d\chi/da)_0$.
 
 As the universe expands according to this solution, $\chi$ carries an
 ever-smaller fraction of the energy density since $K(a)\propto 1/a^6$
 falls faster than does the dominant matter or radiation, $\rho
 \propto 1/a^n$.  Eventually $K$ becomes comparable in size to $V(a)$,
 which for large $a$ asymptotes to a constant value: $V(a) \to
 V_\infty = V_0 e^{-c}$.  At this point the approximation of dropping
 the scalar potential in eq.~\pref{ChiEq} breaks down, and we match
 onto the slow-roll solution discussed above.  The amount of expansion
 which occurs before $V$ and $K$ are comparable is determined by
 $a/a_0 = (e^c K_0/V_0)^{1/6}$.
 
 What is remarkable about this solution is that the constant $c$
 cannot be made arbitrarily large. To see this, notice that $c$ can be
 related to the initial kinetic energy by using $K = \frac12 H^2 a^2
 (d\chi/da)^2$.  Using the upper limit $K \le 3 H^2$, which is
 dictated by the inequality $K \le \rho$ and the Friedman equation, we
 see that $c \le \sqrt{6}\lambda (3 - 1/p)$. For $\lambda = 4$ and $p
 = 1/2$ (radiation domination) we have $c \le 4 \sqrt{6} \sim 9.6$.
 
 This upper limit on $c$ implies $V$ cannot fall by more than a factor
 of $e^{-c} \le e^{-\lambda \sqrt{6}}$ during a radiation dominated
 epoch, despite the fact that $r$'s evolution is being dominated by
 its {\it kinetic} energy. Since $V \propto 1/r^8$ we see that $r/r_0
 < e^{\lambda\sqrt{6}/8} = e^{0.3 \lambda}$ varies hardly at all
 during the entire kinetic roll. Physically the paradoxical upper limit
 on $r$ during a kinetic dominated roll arises due to the Hubble
 damping which increases with the kinetic energy, if the kinetic
 energy gets large. This observation has also been made in another context
 in ref.~\cite{chiba}.
 
 A similar result holds for the kinetic roll in the scalar-dominated
 phase, although with results which differ strongly depending on the
 value of $\lambda$. If $\lambda^2 < 6$, the above analysis applies
 directly, with $p = 2/\lambda^2$. As before $K \propto 1/a^6$ and 
 \begin{equation} 
V(a) = V_0 \exp\left\{ c \left[ \left({a_0 \over
 a}\right)^{3-\lambda^2/2} - 1 \right] \right\}.  
\end{equation} 
 Provided $\lambda^2 < 6$, $V(a)$ again asymptotes to $V_0 e^{-c}$,
 with $c$ once more subject to a very general upper bound. Once $K$
 falls to be of order $V$, this solution matches onto a slow roll
 solution, and approaches the tracker solution, which exists for
 $\lambda^2 < 6$.
 
 If $\lambda^2 > 6$, however, the large-$a$ limit of $V(a)$ is very
 different.  In this case $V(a) \propto \exp[- |c|
 (a/a_0)^{(\lambda^2-6)/2} ]$, implying $V(a)$ falls very quickly to
 zero as the roll proceeds. This solution describes the attractor for
 the case $\lambda^2 > 6$, which is a kinetic-dominated roll rather
 than a tracker solution.
 
 \subsubsection{Realistic Complications}
 
 Although these solutions describe well most features of the numerical
 evolution we describe below, there are also important
 deviations. These deviations arise because $\eta$ and the quantities
 $\lambda$ and $V_p$ (of the potential $V = V_p \;e^{- \lambda \chi}$)
 used in the numerical evolution are all polynomials in
 $\chi$. Motivated by the microphysical discussion given in section II we
 choose: 
\begin{equation} \eta = \eta_0 + \eta_1 \hat\chi; \quad
 \lambda(\chi) = \lambda_0 + \lambda_1 \hat\chi; 
\end{equation} 
 \begin{equation} \hbox{and}\quad V_p = V_0 + V_1 \hat\chi +
 \frac{V_2}{2} \hat\chi^2, 
\end{equation}  
with $\eta_0$, $\lambda_0$
 and $V_0$ all taken to be $O(1)$; $\eta_1, \lambda_1$ and $V_1$
 assumed to be $O(\epsilon)$ and $V_2$ taken $O(\epsilon^2)$, where
 $\epsilon \sim 1/50$. (The second-order quantity $V_2$ is included
 here in order to ensure $V$ does not change sign for any value of
 $\chi$.)
 
 Because these quantities vary more slowly with $\chi$ than does the
 exponential in $V$, their evolution over cosmological timescales is
 also slow, and the above approximate solutions provide a good
 description for most of the universe's history. We have found that
 their $\chi$-dependence can nevertheless play an important role when
 constructing viable cosmologies, by allowing important deviations
 from the approximate solutions, including: 
 \begin{itemize} 
\item  The $\chi$-dependence of $V_p$ becomes important for those $\chi$
 satisfying $V_p'/V_p \approx \lambda$, since the potential $V$ can
 acquire stationary points near such points. These kinds of stationary
 points can slow or trap the cosmological $\chi$ roll, in much the
 same way as was exploited in ref.~\cite{AS1}.  

\item The
 $\chi$-dependence of $\lambda$ can cause transitions between the
 domains of attraction of the various attractor solutions. For
 instance, in a scalar dominated era, the evolution of $\chi$ can
 drive $\lambda$ across the dividing point, $\lambda^2 = 6$, forcing a
 transition between the scaling and kinetic-dominated attractor
 solutions.  

\item The $\chi$-dependence of $\eta$ is crucial for
 evading the bounds on long-range forces during the present epoch,
 because it allows $\eta$ to vanish for some choices of $\chi$.  The
 bounds can be evaded provided the current value of $\chi$ is
 sufficiently close to such a zero of $\eta$. Furthermore, as
 described in detail in ref.~\cite{DamourNordtvedt}, in many
 circumstances $\chi$'s cosmological evolution is naturally attracted
 to the zeros of $\eta$ as the universe expands.  \end{itemize}
 
 Because of the natural hierarchy satisfied by the coefficients of
 these functions, the consequences of $\chi$-dependence generically
 occur for $\chi \sim O(1/\epsilon)$, corresponding to $r$ close to
 its present value.
 
 \subsection{A Realistic Cosmological Evolution}
 
 We next describe a realistic cosmology which exploits some of these
 properties.  We present this model as an existence proof that
 cosmologically interesting evolution can really be found using the
 scalar properties which are suggested by the microphysics. We find
 that the main constraint which governs the construction of such
 solutions is the requirement that $r$ not vary appreciably between
 BBN and the present epoch.
 
 \begin{figure}[h]
\centerline{\hbox{\psfig{file=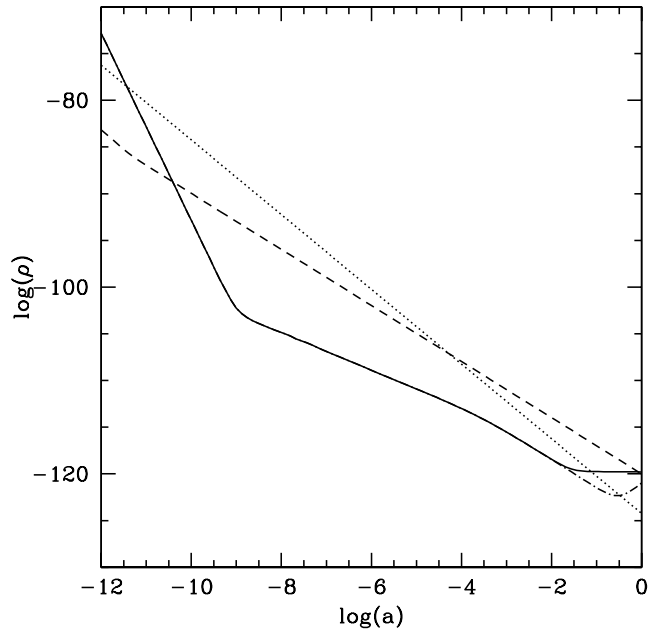,width=3.5in}}}
\caption{The logarithm (base 10) of the energy density (in Planck units)
of various components of the universe,
plotted against the logarithm (base 10) of the Einstein frame
scale factor, $a$ (with $a=1$ at present).
The various curves show the energy density of matter (dashed), 
radiation (dotted) and scalar energy (solid).  The dot-dashed curve
shows the scalar {\em kinetic} energy density, which coincides with
the total scalar density except at late times} 
\label{rho}
\end{figure}

 In order to fix ideas, we imagine starting the universe off at
 $T_*\gsim 1 MeV$, perhaps after an epoch of earlier inflation, in a
 kinetic-dominated roll. We note that such a state is the generic
 endpoint if we start with $\lambda \sim 4 > \sqrt{6}$, as is motivated by the
 tree-level expression for the radion potential, and assume 
the scalar  dominates the energy density of the universe. (As we will
 discuss briefly later, one might chose to make other assumptions
 about the initial conditions.)  

 With this assumption the scalar energy density must eventually fall
 below the density of radiation, thus initiating the
 radiation-dominated era.  We assume that this radiation-domination
 begins shortly before BBN, so that nucleosynthesis occurs during
 radiation domination.
  
 Our second assumption is dictated by the requirement that $r$ does
 not change appreciably between BBN and now. Although the
 kinetic-dominated roll before radiation domination will automatically
 ensure that $V \ll K$, we ask that $V$ be within an order of
 magnitude of to its present value as radiation domination
 begins. This ensures that $r$ will have an initial value at the onset
 of radiation domination which is very close to its present-day
 value. We make this (fairly ugly) assumption to demonstrate the
 feasibility of models along the lines we are presenting, but hope to
 be able to eventually identify a more natural frameworks in which
 such a condition could be ensured by the system's dynamics rather
 than as an initial condition.

 \begin{figure}[h]
 \centerline{\hbox{\psfig{file=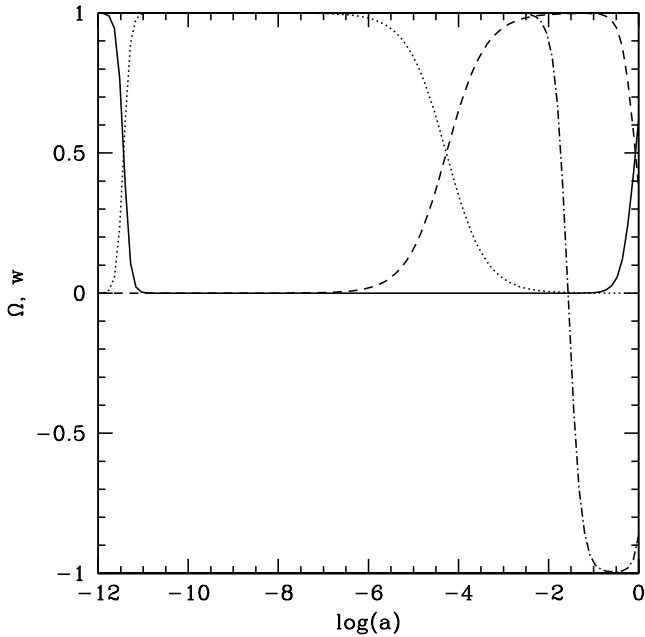,width=3.5in}}}
 \caption{The same plot as Figure \ref{rho}, but with the energy densities 
 given as fractions of the critical density. 
 The various curves show the energy density of matter $\Omega_m$ (dashed), 
 radiation $\Omega_r$ (dotted) and total scalar energy $\Omega_\chi$ (solid),
 while the dash-dot curve gives the equation of state
 parameter, $w = p/\rho$, for the scalar field.}
 \label{omega}
 \end{figure}

 For $\lambda \sim 4$, once the universe becomes radiation dominated
 the scalar in the domain of validity of the tracker solutions
 described above. With the given initial conditions the scalar enters
 radiation domination in a kinetic-dominated roll, and so tries to
 approach a tracker solution by first damping its kinetic energy in a
 continued kinetic-dominated roll, and then performing a slow-roll at
 fixed $r$. We have seen that the universe expands by an amount
 $(a/a_0)^2 \sim K_0/V_0$ during the transient phase, before the
 tracker solution is reached, so given the low values of $V_0$ assumed
 at the onset of radiation domination, $K_0/V_0$ is very large, and so
 the transient evolution to the tracker can easily take the entire
 time from BBN to the present epoch.
 
 We have seen on general grounds that the value of $r$ does not change
 by more than a factor of $O(1)$ during this transient phase,
 including both the transient's kinetic-dominated and slow-rolling
 parts.  The challenge in making this scenario realistic is to have this
 transient evolution survive right up to the present day, without
 first being intercepted by a tracker solution.
 
 \begin{figure}[h]
 \centerline{\hbox{\psfig{file=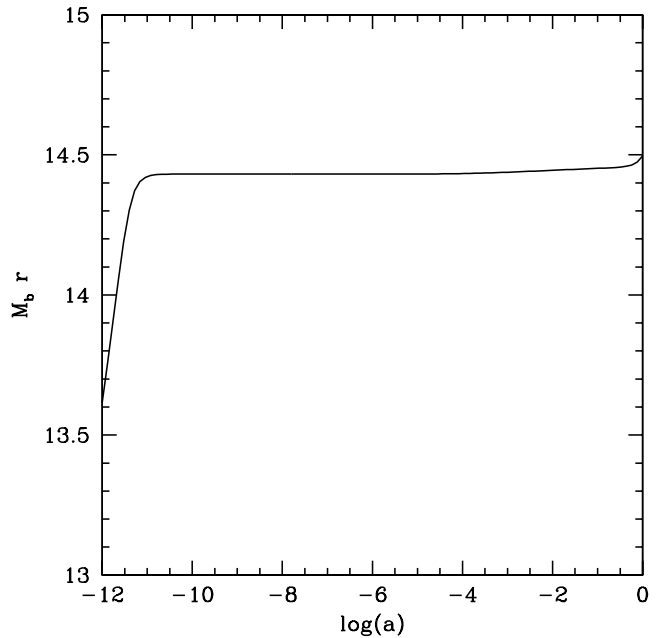,width=3.5in}}}
 \caption{The logarithm (base 10) of the dimensionless quantity $M_b r$,
 plotted against the logarithm (base 10) of the Einstein frame
 scale factor, $a$ (with $a=1$ at present),
 showing how the radion has not evolved appreciably between nucleosynthesis and the
 present epoch.}
 \label{radion}
 \end{figure} 

 We have been able to find initial conditions and plausible couplings
 for which this is accomplished. One such is illustrated in Figures
 (1) through (4), which plot the evolution of various quantities from
 the onset of radiation domination to the present. The evolution
 described by this simulation uses the values suggested by the
 lowest-order microscopic action, $\eta_0 = 1$, $\lambda_0=4$ and
 $V_0=1$, plus the following correction terms: $\eta_1 = -0.015$,
 $\lambda_1 = 0$, $V_1 = -0.030$ and $V_2 = 0.00046$.
 
\begin{figure}[h]
 \centerline{\hbox{\psfig{file=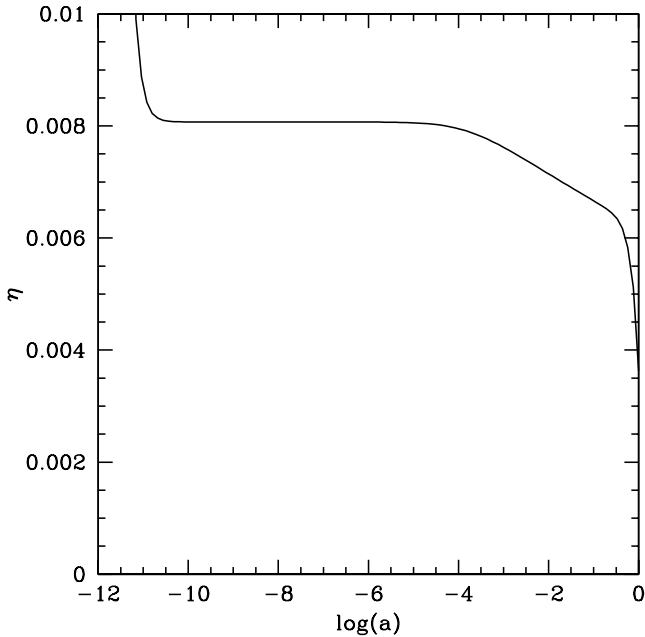,width=3.5in}}}
 \caption{The scalar coupling function, $\eta$,
 plotted against the logarithm (base 10) of the Einstein frame
 scale factor, $a$ (with $a=1$ at present),
 showing how $\eta$ is small enough to satisfy the present-day bounds on the existence
 of long-range scalar forces.}
 \label{alpha}
 \end{figure}

 Figure (\ref{rho}) shows how the energy density in the universe is
 distributed between radiation, matter and the kinetic and potential
 energy of the scalar field, as a function of the universal scale
 factor, $a$, in the Einstein frame. Figure (\ref{omega}) plots the same
 information, but normalized as a fraction of the total universal
 energy density, and also plotting the scalar equation-of-state
 parameter, $w = p_\chi/\rho_\chi$. Figure (\ref{radion}) gives a plot of $r$ vs
 scale factor, showing that $r$ does not vary appreciably between BBN
 and the present epoch. Finally, Figure (\ref{alpha}) shows the evolution of
 $\eta$ against scale factor, showing that $\eta$ evolves to
 sufficiently small values at the present epoch.

 As is evident from the figures, this solution satisfies all of the
 cosmological and present-day bounds we 
 were listed earlier. This was accomplished by choosing appropriately
 the constants governing the scalar interactions, by ensuring $V$ to
 have a minimum relatively near to a zero of $\eta$. This was required
 because although $r$ naturally does not evolve significantly for most
 of the universe's history (after the onset of radiation domination),
 for pure exponential potentials it generically tends to join a
 tracker solution near the onset of matter domination. Because $r$
 evolves too far once in this tracker solution, it must be
 avoided. The solution illustrated in the figure does so by having
 $\chi$ become snagged by the minimum of $V$ before entering into the
 matter-dominated phase.

 \section{Discussion}
 
 We have shown that viable cosmologies may be built using the radion
 as the quintessence field.  We use radion interactions which were
 suggested by a microphysical model which was recently proposed to
 naturally solve the hierarchy problem within a brane-world
 framework. The radion in this model is naturally light enough to play
 a role in late-time cosmology, and this small mass is technically
 natural in that it is not destabilized by radiative corrections. The
 model also satisfies all current bounds which constrain the existence
 of long-range scalar-mediated forces.
 
 The success of the radion in filling the quintessence role is
 somewhat of a surprise, since radion models have long been believed
 to have phenomenologically unacceptable couplings. The model evades
 these problems because the couplings are predicted to be weakly field
 dependent, allowing the scalar couplings to evolve over cosmological
 timescales. The model evades all present-day bounds on new forces
 because the relevant couplings evolve towards small values at late
 times.
 
 Although most of the parameters and initial conditions chosen were
 natural in size, the model does have two features which we believe
 need further improvement.  First, although all scalar couplings were
 chosen with natural sizes, the precise values chosen were adjusted to
 arrange the scalar potential to have a minimum close to a zero of
 $\eta$. We would prefer to find an attractor solution, \`a la Damour
 and Nordtvedt \cite{DamourNordtvedt}, which more naturally draws
 $\eta$ to small values in the present epoch.
 
 A second unsatisfactory feature of the cosmology presented was its
 reliance on the entering of radiation domination with $r$ having
 close to its present value.  Although large values of $r$ are
 naturally generated by kinetic-dominated rolls during scalar
 dominated epochs, there is no natural reason why they should have
 precisely the current value.  This problem might be resolved by a
 more developed understanding of radion initial conditions. Perhaps
 one could argue that there are no a-priori constraints on the
 initial conditions for $r$.  In that picture, it may be the case that
 the largest amount of phase space that matches the observation that
 Newton's constant is constant and non-zero today would be one where
 $r$ has the current value throughout the cosmic evolution. (In that
 reasoning, the constancy of $G$ is not a prediction, but an input.)
 
 A natural way to alleviate both of these problems would be to
 consider more moduli than just the radion, since other moduli would
 be less constrained by the requirement that Newton's constant not
 appreciably change since nucleosynthesis. We believe work along this
 lines to be worthwhile, given the model's successful addressing of
 the naturalness issues of quintessence models, and the promising
 cosmology to which it leads despite its being devised to solve purely
 microphysical problems such as the hierarchy problem.

 \section{Acknowledgments}
 
 We would like to acknowledge useful conversations with A. Aguirre, S.Basu  
 and S. Gratton, and the Aspen Center for Physics for providing the
 pleasant environs in which this work started. CB's research is
 partially funded by NSERC (Canada), FCAR (Qu\'ebec) and the Ambrose
 Monell Foundation. AA and CS are supported by DOE grant
 DE-FG03-91ER40674 and U.C. Davis.

\end{document}